\documentclass[aps,superscriptaddress,amsmath,amssymb,twocolumn,showpacs,floatfix,english,noshowpacs]{revtex4-2}

\usepackage{url}
\usepackage{bm,bbm}
\usepackage{graphicx}
\usepackage{xcolor}

\usepackage[colorlinks=true, urlcolor=blue, linkcolor=blue, citecolor=blue, pdftex]{hyperref}
\usepackage[english]{babel}

\usepackage{soul}
\usepackage{blindtext}
\usepackage{lipsum}
\usepackage{nicefrac}
\usepackage{cancel}

\newcommand{\dagga}{{\phantom{\dagger}}}

\begin{document}

\title{Interaction-induced phases in the half-filled Bernevig-Hughes-Zhang model in one dimension}
\author{Roberta Favata}
\email{roberta.favata@phd.units.it}
\affiliation{Dipartimento di Fisica, Universit\`a di Trieste, Strada Costiera 11, I-34151 Trieste, Italy}
\author{Davide Piccioni}
\affiliation{International School for Advanced Studies (SISSA), Via Bonomea 265, I-34136 Trieste, Italy}
\author{Alberto Parola}
\affiliation{Dipartimento di Scienza e Alta Tecnologia, Universit\`a dell’Insubria, Como, Italy}
\author{Federico Becca}
\affiliation{Dipartimento di Fisica, Universit\`a di Trieste, Strada Costiera 11, I-34151 Trieste, Italy}

\date{\today}

\begin{abstract}
We explore the ground-state properties of a one-dimensional model with two orbitals per site, where, in addition to atomic energies $\pm M$, intra- and inter-orbital 
hoppings, the intra-orbital Hubbard ($U$) and nearest-neighbor density-density ($V$) repulsions are included. Our results are primarily based on a Jastrow-Slater
wave function and variational Monte Carlo methods but also corroborated by density-matrix renormalization group calculations. In the non-interacting limit, when 
varying $M>0$, a gapless point separates a trivial phase from a topological one. The inclusion of a finite Hubbard-$U$ repulsion does not give rise to any phase 
transition within the topological region, inducing a smooth crossover into a Haldane (spin gapped) insulator; notably, the string-order parameter, which characterizes 
the latter phase, is already finite in the non-interacting limit. Most importantly, at finite values of $U$, the transition between the trivial and topological states 
is not direct, since an emergent insulator, which shows evidence of sustaining gapless spin excitations, intrudes between them. A small $V$ interaction further 
stabilizes the intermediate insulator, while a sufficiently large value of this nearest-neighbor repulsion gives rise to two different charge-density wave insulators, 
one fully gapped and another still supporting gapless spin excitations. Our results demonstrate the richness of multi-orbital Hubbard models, in the presence of a 
topologically non-trivial band structure, and serve as a basis for future investigations on similar two-dimensional models.
\end{abstract}

\maketitle

\section{Introduction}

The electron-electron interaction may have strong effects in one or two spatial dimensions, leading to a variety of phenomena that cannot be captured within the 
independent-electron picture; well-established examples are Tomonaga-Luttinger~\cite{haldane1981} or Luther-Emery~\cite{luther1974} liquids in one dimension and 
Mott insulators~\cite{mott1968} or the fractional quantum Hall effect~\cite{laughlin1983} in two dimensions; more controversial cases are given by high-temperature 
superconductors, where the origin of the pseudogap phase is still under debate~\cite{fradkin2015}. Recently, a boost in the field has been given by the discovery 
of Moir\'e materials (e.g., twisted graphene superlattices)~\cite{cao2018}, which revealed the presence of a wide variety of phases when varying the filling factor 
of the Moir\'e bands~\cite{lu2019}. These are examples in which strong electron correlation exists on top of a band structure that features topological properties, 
thus giving general motivation to understand the fate of topological insulators beyond the band theory (e.g., when the Coulomb repulsion is included). The most 
challenging problem would be to assess the possibility that unconventional superconductivity (i.e., not mediated by phonons) may emerge from a correlated topological 
insulator when the electron filling is varied. A significant but more focused goal is to examine the sole impact of electron-electron interactions on a topological 
insulator, maintaining a fixed electron density. In this framework, it is of great interest to study the nature of phase transitions between trivial, topological, 
and Mott insulators, potentially involving additional phases that may intrude on them. In fact, in the case of non-interacting fermions, the transition between 
a topological insulator and a trivial gapped phase is generally continuous, characterized by a single semi-metallic point at which the band gap closes (provided 
the symmetries of the system are preserved)~\cite{haldane1988,kane2005,hasan2010}. Recently, a few investigations have focused on the role of on-site interactions 
in topological insulators, e.g., Kane-Mele~\cite{rachel2010,hohenadler2011,yu2011,yamaji2011,hohenadler2012,budich2012,hohenadler2014} or Bernevig-Hughes-Zhang 
(BHZ) models~\cite{tada2012,tsuneys2013,budich2013,amaricci2015}. In particular, it has been noted that the quantum phase transition between trivial and topological 
phases might become, for sufficiently large interactions, first-order~\cite{budich2013,amaricci2015,imriska2016}. 

While most studies focus on two- and three-dimensional models, one-dimensional (1D) systems are gaining increasing attention in recent years. Indeed, despite their 
reduced dimensionality, 1D systems exhibit a variety of topological phases remarkably similar to those found in higher-dimensional systems. This makes them a 
valuable starting point for studying topological properties, as the reduced complexity of 1D systems simplifies analysis. In addition, 1D topological systems can 
be experimentally realized, by optical lattices~\cite{atala2013,verbin2013} or nanowire materials~\cite{jin2020,liu2022}. From the theoretical side, very accurate 
numerical techniques have been developed in the past, including density-matrix renormalization group (DMRG)~\cite{white1992} and variational quantum Monte Carlo, 
based on Jastrow-Slater wave functions~\cite{beccabook}. While the former one may give numerically exact results, especially within gapped phases, the latter 
approach has been highly improved in the past 20 years (thanks to efficient optimization methods~\cite{sorella2005}), thus allowing us to reach the correct 
description in a variety of situations, including the case of gapless phases~\cite{capello2005,capello2005b}. Importantly, variational wave functions and quantum 
Monte Carlo methods may be easily generalized in two or three dimensions, still keeping a sufficient accuracy to address important aspects of strongly-correlated 
systems.

A few works focused attention on the role of electron-electron interactions on top of 1D Su-Schrieffer-Heeger~\cite{manmana2012,barbiero2018} and 
BHZ~\cite{guo2011,barbarino2019} models by using exact diagonalizations or DMRG methods. In particular, Ref.~\cite{barbarino2019} considered the role of a 
nearest-neighbor Coulomb term, which favors a charge-density-wave (CDW) insulator, showing that the transition between the topological insulator and the CDW one 
becomes first-order in presence of a sufficiently large on-site (intra-orbital) interaction. Still, only a small portion of the phase diagram was investigated 
and many aspects require a more detailed discussion. Some of these aspects are important for future studies of two-dimensional systems but others are important 
{\it per se}, showing the abundance and variety of phases that exist in 1D systems. 

%%%%%%%%%%%%%%%%%%%%%%%%%%%%%%%%%%%%%%%%%%%%%%%%%%%%%%%%%%%%%%%%%%%%%%%%%%%%%%%%%%%%%%%%%%%%%%%%%%%%%%%%%%%%%%%%%%%%%%%%%%%%%%%%%%%
\begin{figure}
\includegraphics[width=\columnwidth]{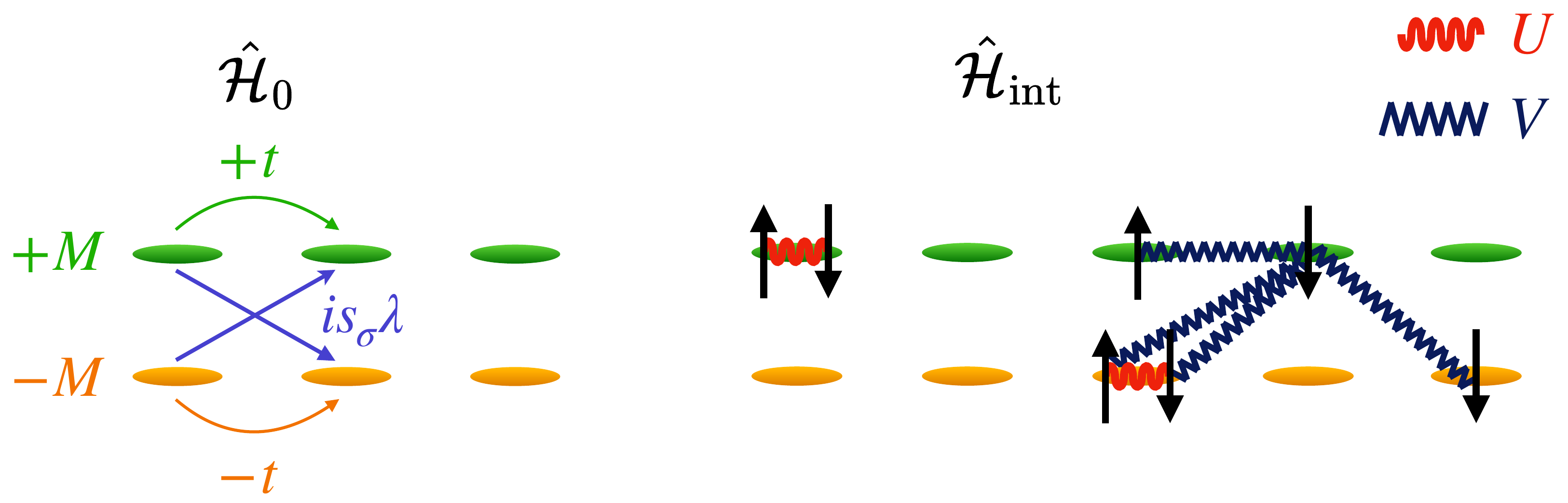}
\caption{\label{fig:lattice}
Illustration of the terms defining the Hamiltonian of Eq.~\eqref{eq:hamilt}. The two orbitals on each site are represented by yellow and green ellipses; on the 
left part, on-site energies ($\pm M$), intra-orbital ($\pm t$), and inter-orbital ($i s_{\sigma} \lambda$) hoppings are depicted; on the right part, intra-orbital 
($U$) and nearest-neighbor ($V$) interactions are shown.}
\end{figure}
%%%%%%%%%%%%%%%%%%%%%%%%%%%%%%%%%%%%%%%%%%%%%%%%%%%%%%%%%%%%%%%%%%%%%%%%%%%%%%%%%%%%%%%%%%%%%%%%%%%%%%%%%%%%%%%%%%%%%%%%%%%%%%%%%%%

Therefore, in this work, we consider the fermionic model with two orbitals per site defined in Refs.~\cite{guo2011,barbarino2019}, whose Hamiltonian corresponds to 
the 1D version of the BHZ model~\cite{bernevig2006}:
\begin{equation}\label{eq:hamilt}
\hat{\cal H} = \hat{\cal H}_{0} + \hat{\cal H}_{\rm int};
\end{equation}
the non-interacting part is given by
\begin{eqnarray}
\hat{\cal H}_{0} &=& \sum_{j=1}^{L} \sum_{\eta,\sigma} \bigl[ M \, \eta \, \hat{n}_{j,\eta,\sigma} 
+ (t \, \eta \, \hat{c}^\dag_{j+1,\eta,\sigma} \hat{c}^\dagga_{j,\eta,\sigma} \nonumber \\
&+& i s_{\sigma} \lambda \, \hat{c}^\dag_{j+1,\eta,\sigma} \hat{c}^\dagga_{j,-\eta,\sigma} + \mathrm{h.c.}) \bigr],
\label{eq:h0}
\end{eqnarray}
where $\hat{c}^\dag_{j,\eta,\sigma}$ ($\hat{c}^\dagga_{j,\eta,\sigma}$) is the creation (annihilation) operator of an electron on the orbital $\eta=\pm 1$ and 
spin $\sigma$ at site $j$, $\hat{n}_{j,\eta,\sigma} = \hat{c}^\dag_{j,\eta,\sigma} \hat{c}^\dagga_{j,\eta,\sigma}$ is the local density (per orbital and spin), 
and $s_{\uparrow}=+1$ or $s_{\downarrow}=-1$. In addition, the density per orbital is defined by $\hat{n}_{j,\eta} = \sum_{\sigma} \hat{n}_{j,\eta,\sigma}$, while 
the total density on site $j$ is $\hat{n}_{j}=\sum_{\eta,\sigma} \hat{n}_{j,\eta,\sigma}$. Following Ref.~\cite{barbarino2019}, the interacting part contains 
intra-orbital on-site interaction and nearest-neighbor density-density repulsion:
\begin{equation}
\hat{\cal H}_{\rm int} = U \sum_{j=1}^{L} \sum_{\eta} \hat{n}_{j,\eta,\uparrow} \hat{n}_{j,\eta,\downarrow} + V \sum_{j=1}^{L} \hat{n}_{j} \hat{n}_{j+1}.
\end{equation}
In the following, we consider a chain with $L$ sites and periodic-boundary conditions at half filling (i.e., fixing the number of total electrons to be $N_e=2L$),
with $M \ge 0$ and $t>0$. The various terms that define the Hamiltonian~\eqref{eq:hamilt} are schematically sketched in Fig.~\ref{fig:lattice}.

The main results of the present study can be summarized as follows. (i) The inclusion of a finite $U$ on top of the topological insulator gives rise to the Haldane 
state (where each orbital is singly occupied and the low-energy sector is described by a spin $S=1$ chain); notably, the string-order parameter~\cite{dennijs1989} 
is already finite for $U=0$. (ii) An insulating phase with {\it gapless} spin excitations intrudes between trivial and topological phases at finite values of $U$; 
this intermediate phase is particularly interesting since it cannot be derived within a simple strong-coupling limit. (iii) The presence of a nearest-neighbor 
interaction $V$ induces density disproportionations: when $V$ dominates all the other energy scales, a fully gapped CDW is obtained, adiabatically connected to 
the atomic limit in which sites with 4 electrons alternate with empty sites (this phase is dubbed CDW$_{4-0}$); in addition, in a region around $V/U \approx 1/4$ 
and $M/U \approx 1/2$, a different CDW exists and can be adiabatically connected to the atomic limit in which sites with 3 electrons alternate with singly-occupied 
sites; here, the spin degrees of freedom are governed by an effective spin $S=1/2$ Heisenberg Hamiltonian, thus leading to {\it gapless} spin excitations (this 
phase is dubbed CDW$_{3-1}$). 

These results are demonstrated by using Jastrow-Slater wave functions and variational Monte Carlo techniques~\cite{beccabook}; furthermore, DMRG calculations on 
the same clusters, with periodic boundary conditions in the Hamiltonian (but using an {\it Ansatz} that is defined on open-boundary conditions) have been performed,
corroborating the Monte Carlo results.

The outline of the paper is as follows: in section~\ref{sec:atomic}, we list the symmetries of the BHZ model, summarize its non-interacting properties, and discuss 
the ``atomic'' limit, obtained by taking $t=\lambda=0$, which already contains most of the phases that are stabilized in the general case, with $t \ne 0$ and 
$\lambda \ne 0$; in section~\ref{sec:wavefun}, we discuss the form of the variational wave function that is used in this work; then in section~\ref{sec:results}, 
we show the main numerical results for the interacting 1D BHZ model; finally, in section~\ref{sec:concl}, we draw our conclusions.

\section{Preliminary considerations on the 1D BHZ model}\label{sec:atomic}

\subsection{The symmetries}

The model of Eq.~\eqref{eq:hamilt} is invariant under a few local transformations. The first one is the (anti-unitary) time reversal (TR) symmetry, defined by:
\begin{eqnarray}
&& {\cal T}_R \, \hat{c}^\dag_{j,\eta,\uparrow} \, {\cal T}^{-1}_R = +\hat{c}^\dag_{j,\eta,\downarrow}, \\
&& {\cal T}_R \, \hat{c}^\dag_{j,\eta,\downarrow} \, {\cal T}^{-1}_R = -\hat{c}^\dag_{j,\eta,\uparrow}.
\end{eqnarray}
The second one is a (unitary) spin-flip, with a sign depending on the orbital (SF+$\eta$ sign):
\begin{eqnarray}
&& {\cal F} \, \hat{c}^\dag_{j,\eta,\uparrow} \, {\cal F}^{-1} = \eta \hat{c}^\dag_{j,\eta,\downarrow}, \\
&& {\cal F} \, \hat{c}^\dag_{j,\eta,\downarrow} \, {\cal F}^{-1} = \eta \hat{c}^\dag_{j,\eta,\uparrow}.
\end{eqnarray}
At half filling, the Hamiltonian is also invariant under a (unitary) particle-hole transformation, with orbital flip ($\eta$-PH):
\begin{equation}\label{eq:PH}
{\cal P} \, \hat{c}^\dag_{j,\eta,\sigma} \, {\cal P}^{-1} = \hat{c}^{\dagga}_{j,-\eta,\sigma}.
\end{equation}
In addition, the Hamiltonian~\eqref{eq:hamilt} is invariant under a $U(1)$ spin symmetry:
\begin{equation}\label{eq:U1}
{\cal U} \, \hat{c}^\dag_{j,\eta,\sigma} \, {\cal U}^{-1} = e^{i\theta} \hat{c}^\dag_{j,\eta,\sigma},
\end{equation}
and lattice symmetries, notably translation (T) and inversion with a sign depending on the orbital (I+$\eta$ sign):
\begin{eqnarray}\label{eq:trans}
&& {\cal T} \, \hat{c}^\dag_{j,\eta,\sigma} \, {\cal T}^{-1} = \hat{c}^\dag_{j+1,\eta,\sigma}, \\
&& {\cal I} \, \hat{c}^\dag_{j,\eta,\sigma} \, {\cal I}^{-1} = \eta \hat{c}^\dag_{L+2-j,\eta,\sigma}.
\label{eq:Inv}
\end{eqnarray}

%%%%%%%%%%%%%%%%%%%%%%%%%%%%%%%%%%%%%%%%%%%%%%%%%%%%%%%%%%%%%%%%%%%%%%%%%%%%%%%%%%%%%%%%%%%%%%%%%%%%%%%%%%%%%%%%%%%%%%%%%%%%%%%%%%%
\begin{figure}
\includegraphics[width=\columnwidth]{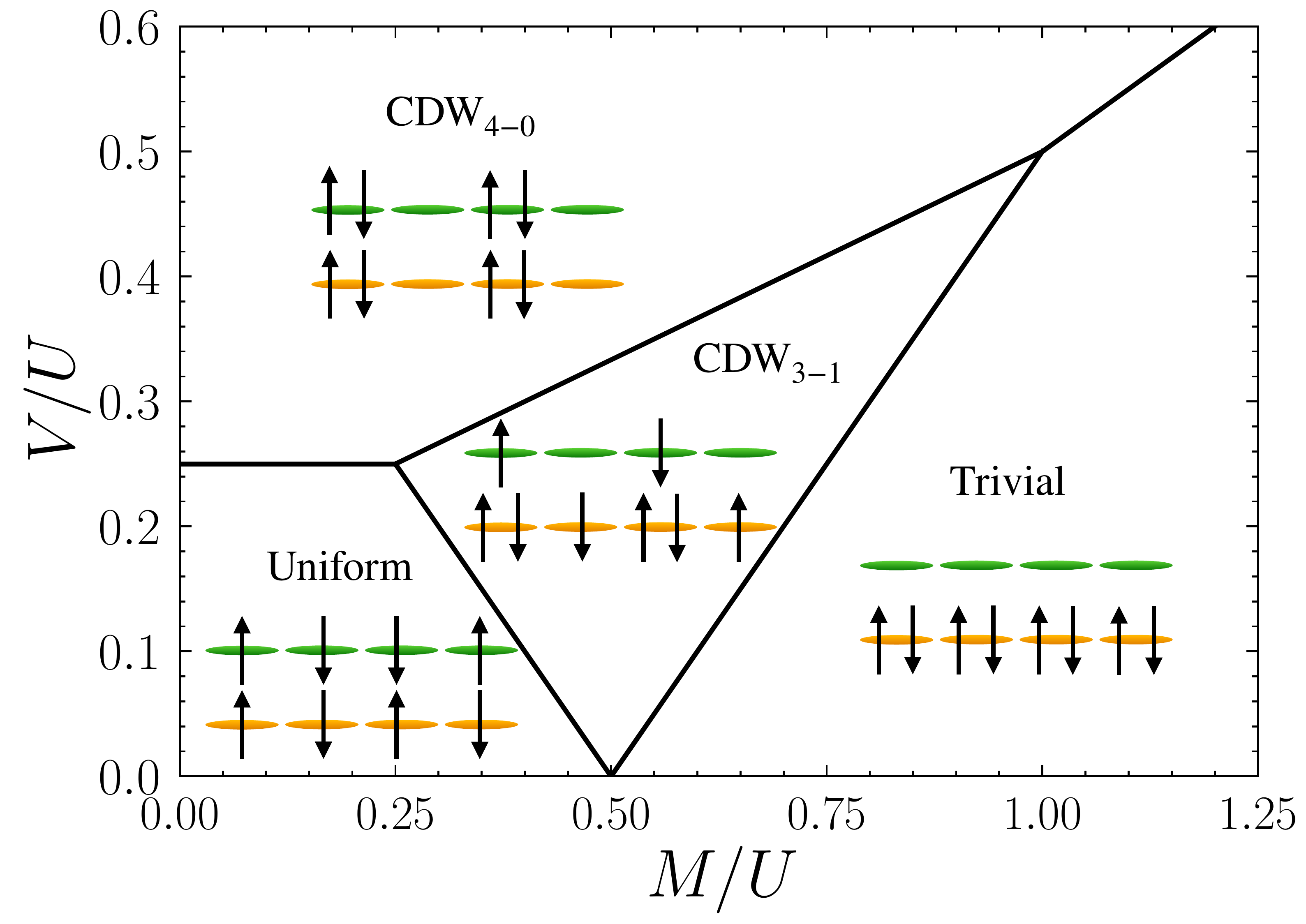}
\caption{\label{fig:atomic}
Phase diagram in the atomic limit, with $t=\lambda=0$. The Uniform and Trivial phases have two electrons on each site; in the former, electrons occupy different 
orbitals (with an arbitrary spin configuration), while in the latter they reside in the same orbital (the one with the lowest on-site energy). The other two phases 
(CDW$_{4-0}$ and CDW$_{3-1}$) have non-uniform density patterns; within the CDW$_{4-0}$, sites with 4 electrons alternate with empty sites, and in the CDW$_{3-1}$ 
there is an alternation between sites with 1 and 3 electrons (and an arbitrary spin configuration). In this case, all transition lines are first-order.}
\end{figure}
%%%%%%%%%%%%%%%%%%%%%%%%%%%%%%%%%%%%%%%%%%%%%%%%%%%%%%%%%%%%%%%%%%%%%%%%%%%%%%%%%%%%%%%%%%%%%%%%%%%%%%%%%%%%%%%%%%%%%%%%%%%%%%%%%%%

\subsection{The non-interacting limit}

Here, we briefly summarize the properties of the non-interacting limit of the 1D BHZ model~\cite{barbarino2019} (always assuming $t>0$ and $M \ge 0$). In presence 
of periodic-boundary conditions, the model~\eqref{eq:h0} can be diagonalized by a Fourier transformation, where the Block Hamiltonian is written in terms of Pauli 
matrices in the spin $\boldsymbol{\sigma}$ and orbital $\boldsymbol{\tau}$ spaces:
\begin{equation}
H_0(k) = [2t \cos(k) + M] \tau_z \otimes \sigma_0 + 2 \lambda \sin(k) \tau_x \otimes \sigma_z,
\end{equation}
where $\sigma_0$ denotes the $2\times 2$ identity matrix. The Hamiltonian consists of two diagonal blocks in spin space, leading to two spin-degenerate copies of 
the bands $E_k= \pm \sqrt{[2t \cos(k) + M]^2 + [2 \lambda \sin(k)]^2}$. At half filling (e.g., 2 electrons per site on average), trivial and topological phases are 
obtained for $M/t>2$ and $M/t<2$, respectively. These two cases can be distinguished by the spin-resolved Zak phase $\phi_{\sigma}$ (the time-reversal symmetry 
implies that the full Zak phase is always vanishing), as obtained by the eigenstates of either the spin up or down blocks of $H_0(k)$~\cite{zak1989}. 
Then, $\phi_{\sigma}=0$ characterizes the trivial region, while the topological phase has $\phi_{\sigma}=\pi$.

We mention that, in the presence of open-boundary conditions, edge states appear in the topological phase. In close analogy with the Su-Schrieffer-Heeger model,
four zero-energy states (two for each spin value) are present for $M/t<2$. This is exact on any cluster taking $M=0$ and $\lambda=t$. At half filling, these four levels
must be filled with 2 electrons, leading to six many-body wave functions. We remark that two of them have both electrons on the same edge, while the other four have the two electrons on opposite edges.

\subsection{The ``atomic'' limit}

It is instructive to consider the ``atomic'' limit of the 1D BHZ model, i.e., the case with no hopping terms ($t=\lambda=0$). The phase diagram for the half-filled 
case can be easily worked out; see Fig.~\ref{fig:atomic}. Here four distinct states are obtained: two  of them have uniform density (i.e., two electrons on each 
site), while the other two have density disproportionations. For moderate values of $V$ and large on-site energy $M$, a trivial insulator is stabilized, with two 
electrons in the same orbital (the one having a negative on-site energy). When $M$ is small, the ground state has one electron on each orbital, thus having an 
exponentially large degeneracy (e.g., $2^{2L}$) due to the spin component of each electron. Between these two cases, the CDW$_{3-1}$ state is stabilized; also 
here a large degeneracy is present (e.g., $2 \times 2^{L}$), due to the singly occupied orbitals and the two-fold degeneracy due to translations. Finally, for 
large $V$ interactions, the CDW$_{4-0}$ state becomes the lowest-energy configuration; this state is fully gapped, while being two-fold degenerate due to the 
breaking of translation symmetry.

In order to fully specify the strong-coupling phase diagram, we have to assess the spin state in the two regions (Uniform and CDW$_{3-1}$) where an exponential 
degeneracy is present at $t=\lambda=0$. This goal can be accomplished by using second-order perturbation theory in the hopping amplitudes $t$ and $\lambda$, giving 
rise to Heisenberg-like models, written in terms of the spin operators:
\begin{equation}\label{eq:spin}
\hat{S}^{\alpha}_{j,\eta} = \frac{1}{2} \sum_{\sigma,\sigma^\prime} \hat{c}^\dag_{j,\eta,\sigma} \tau^{\alpha}_{\sigma,\sigma^\prime}
\hat{c}^\dagga_{j,\eta,\sigma^\prime}
\end{equation}
where $\tau^{\alpha}$ indicates one of the three Pauli matrices for $\alpha=x$, $y$, and $z$.

Let us begin with the CDW$_{3-1}$ phase, defined by the inequalities $|U-2M|<2V$ and $U+2M>6V$, where each site $j$ hosts a spin $S=1/2$. The ground-state energy per 
site at $t=\lambda=0$ is $E_0 = 3V+\frac{U}{2}-M$. Virtual hoppings of amplitude $t$, which do not modify the global orbital occupancy, are always possible, giving 
a constant contribution to the energy per site $\Delta E_{t}=-\frac{2t^2}{3V}$. Analogously an inter-orbital hopping of amplitude $is_\sigma\lambda$ from a doubly 
occupied site to an empty site gives a constant energy per site $\Delta E_{\lambda}=-\frac{2\lambda^2}{3V-U+2M}$. Instead, the inter-orbital hopping of the same 
amplitude $is_\sigma\lambda$ between singly occupied sites favors antiferromagnetic ordering and gives a term in the effective spin Hamiltonian reproducing the usual
spin-$1/2$ Heisenberg model:
\begin{equation}\label{eq:hamIII}
{\cal H}_{CDW_{3-1}}=J_{3-1} \sum_{j=1}^L \left [\hat{\bf S}_j \cdot \hat{\bf S}_{j+1} -\frac{1}{4}\right ],
\end{equation}
where $\hat{\bf S}_{j}=\hat{\bf S}_{j,+} + \hat{\bf S}_{j,-}$ is the total spin operator on the $j$ site and
\begin{equation}
J_{3-1} = \frac{2\lambda^2}{U-2M+3V} + \frac{2\lambda^2}{U+2M-5V}.
\end{equation}
This form of the effective Hamiltonian justifies the presence of gapless spin excitations in CDW$_{3-1}$ phase. 

The case of the Uniform region is different, since there are two $S=1/2$ spins on different orbitals. The energy per site at $t=\lambda=0$ is $E_0=4V$ and the two 
inequalities $U>4V$ and $U>2V+2M$ hold in this phase. Every virtual hopping between neighboring sites favors antiferromagnetic ordering both on the same orbital 
and on different orbitals giving an effective Hamiltonian of the form
\begin{eqnarray}\label{eq:hamI}
{\cal H}_{U}&=& J\,\sum_{j=1}^L \sum_\eta \left [\hat{\bf S}_{j,\eta} \cdot \hat{\bf S}_{j+1,\eta} -\frac{1}{4}\right ]  \nonumber\\
	    &+& J^\prime\,\sum_{j=1}^L \sum_\eta \left [\hat{\bf S}_{j,\eta} \cdot \hat{\bf S}_{j+1,-\eta} -\frac{1}{4}\right ]
\end{eqnarray}
with 
\begin{eqnarray}
J &=& \frac{4t^2}{U-V}, \\
J^\prime &=& \frac{2\lambda^2}{U-V-2M} + \frac{2\lambda^2}{U-V+2M}.
\end{eqnarray}
The ground state of this model can be easily figured out in the classical limit, where  the alignment of the two spins on different orbitals at the same site is 
energetically favored, suggesting the picture of an effective spin-$1$ model governing the low-energy behavior of the system. This conclusion can be made rigorous 
along the special line $\lambda=t\sqrt{1-\frac{4M^2}{(U-V)^2}}$, following the argument detailed in Ref.~\cite{white1996}. In this case, the amplitudes of the two 
terms in Eq.~\eqref{eq:hamI} coincide, the total spin on each site $\hat{\bf S}_{j}$ commutes with ${\cal H}_U$ and the ground state belongs to the subspace where 
the two spins form a triplet on each site. This implies that, in the low-energy sector, the Hamiltonian ${\cal H}_U$ maps into a spin-$1$ Heisenberg chain, whose 
ground state is known to exhibit string order and gapped excitation spectrum (Haldane phase). 

In the following, we take $t=\lambda=1$ and show how these four states evolve by varying $M$, $U$, and $V$. 

\section{Variational wave functions}\label{sec:wavefun}

In this work, we use a Jastrow-Slater wave function defined as
\begin{equation}\label{eq:wf}
|\Psi_{\rm var} \rangle = \mathcal{J}_{\rm s} \mathcal{J}_{\rm d} \mathcal{P}_{\{S^{z}=0\}}|\Phi_{0} \rangle,
\end{equation}
where $\mathcal{J}_{s}$ and $\mathcal{J}_{d}$ are spin-spin and density-density Jastrow factors:
\begin{eqnarray}\label{eq:spinjas}
\mathcal{J}_{\rm s} &=& \exp \left (\frac{1}{2} \sum_{i,j} \sum_{\alpha,\beta} u_{i,j}^{\alpha,\beta} \hat{S}^{z}_{i,\alpha} \hat{S}^{z}_{j,\beta} \right ), \\
\mathcal{J}_{\rm d} &=& \exp \left (\frac{1}{2} \sum_{i,j} \sum_{\alpha,\beta} v_{i,j}^{\alpha,\beta} \hat{n}_{i,\alpha} \hat{n}_{j,\beta} \right ), 
\label{eq:densjas}
\end{eqnarray}
where the pseudo-potentials $u_{i,j}^{\alpha,\beta}$ and $v_{i,j}^{\alpha,\beta}$ are variational parameters, which depend on the spatial distance $|i-j|$;
in addition, we set $u_{i,j}^{\alpha,\beta}=u_{i,j}^{\beta,\alpha}$, and $v_{i,j}^{\alpha,\beta}=v_{i,j}^{\beta,\alpha}$ for $\alpha \ne \beta$ and fix the 
intra-orbital terms to be the same for the two orbitals. Furthermore, $\mathcal{P}_{\{S^{z}=0\}}$ is the projection onto the subspace with 
$S^{z}_{\rm tot}= \sum_{i,\alpha} \hat{S}^{z}_{i,\alpha}=0$.

The Jastrow factors induce correlation effects onto the non-interacting Slater determinant $|\Phi_{0}\rangle$, which is obtained as the ground state of the auxiliary 
(quadratic) Hamiltonian, which contains three terms:
\begin{equation}\label{eq:auxiliary}
\hat{\cal H}_{\rm aux} = \hat{\cal H}_{\rm band} + \hat{\cal H}_{\rm CDW} + \hat{\cal H}_{\rm AF}.
\end{equation}
The first one contains  the ``mass'' term and the intra- and inter-orbital hoppings (including an on-site inter-orbital one):
\begin{eqnarray}
\hat{\cal H}_{\rm band} &=& \sum_{j=1}^{L} \sum_{\eta,\sigma} \bigl[ \tilde{M} \, \eta \, \hat{n}_{j,\eta,\sigma} 
+ \tilde{\lambda}_{0,\sigma} \hat{c}^\dag_{j,\eta,\sigma} \hat{c}^\dagga_{j,-\eta,\sigma} \nonumber \\
&+& \sum_{k=1,2} (\tilde{t}_{k} \, \eta \hat{c}^\dag_{j+k,\eta,\sigma} \hat{c}^\dagga_{j,\eta,\sigma} + \mathrm{h.c.}) \nonumber \\
&+& (\tilde{\lambda}_{k,\eta,\sigma} \, \hat{c}^\dag_{j+k,\eta,\sigma} \hat{c}^\dagga_{j,-\eta,\sigma} + \mathrm{h.c.}) \bigr];
\end{eqnarray}
here $\tilde{M}$, $\tilde{\lambda}_{0,\sigma}$, $\tilde{t}_{k}$ (with $\tilde{t}_{1} \equiv t$, fixed to define the energy scale), and 
$\tilde{\lambda}_{k,\eta,\sigma}$ are variational parameters, which allow a renormalization of the band structure. Most importantly, the presence of non-vanishing 
inter-orbital hopping parameters may break some of the symmetries discussed above (TR, SF+$\eta$ sign, $\eta$-PH, and I+$\eta$ sign, while translational symmetry
is always preserved); see Appendix~\ref{sec:app} for details. 

The last two terms in Eq.~\eqref{eq:auxiliary} are given by
\begin{eqnarray}
\hat{\cal H}_{\rm CDW} &=& \Delta \sum_{j=1}^{L} (-1)^j \, \hat{n}_{j},
\label{eq:hamCDW}\\
\hat{\cal H}_{\rm AF} &=& h \sum_{j=1}^{L} (-1)^j \sum_{\eta} \eta ( \hat{c}^\dag_{j,\eta,\uparrow} \hat{c}^\dagga_{j,\eta,\downarrow}
+ \mathrm{h.c.}),
\label{eq:hamAF}
\end{eqnarray}
where $\Delta$ and $h$ are additional variational parameters. In particular, the former one is important to induce CDW order in the non-interacting wave function; 
by contrast, the (fictitious) magnetic field $h$ is useful to obtain an accurate {\it Ansatz} in the large-$U$ limit, recovering the results obtained within the 
$S=1$ Heisenberg model~\cite{piccioni2023}. 

The optimization of the variational parameters and the evaluation of the physical properties (e.g., variational energy and correlation functions) can be easily
performed by using Markov chains with simple Metropolis algorithm~\cite{beccabook}. We remark that, in presence of translational symmetry, only a reduced number 
of variational parameters is defined [e.g., $O(1)$ in the Slater determinant and $O(L)$ in the Jastrow factors], allowing for efficient Monte Carlo optimization.
A much more demanding task would be required in presence of open-boundary conditions, which would require a much larger number of variational parameters [e.g.,
$O(L)$ in the Slater determinant and $O(L^2)$ in the Jastrow factors]. Therefore, the present variational approach is not particularly suited to assess the 
presence of edge states in the topological phase. 

Finally, our variational Monte Carlo results are compared to DMRG calculations in which we take a matrix-product state (which has no translational symmetry, although 
the Hamiltonian is always defined with periodic-boundary conditions) and optimize it by means of the algorithm implemented in the ITensor library~\cite{itensor} up 
to a bond dimension $\chi=1600$; the accuracy of the calculations is verified by evaluating the variance of the total energy, which is always below $0.0025t^2$.

\section{Results}\label{sec:results}

Here we present the numerical results, which are obtained through a variational Monte Carlo approach on the wave function described in section~\ref{sec:wavefun}. 
We start from the case with $V=0$, showing how the intra-orbital Hubbard-$U$ interaction modifies the non-interacting trivial and topological phases. Then, 
we consider the effect of the nearest-neighbor repulsion $V$. 

%%%%%%%%%%%%%%%%%%%%%%%%%%%%%%%%%%%%%%%%%%%%%%%%%%%%%%%%%%%%%%%%%%%%%%%%%%%%%%%%%%%%%%%%%%%%%%%%%%%%%%%%%%%%%%%%%%%%%%%%%%%%%%%%%%%
\begin{figure}
\includegraphics[width=\columnwidth]{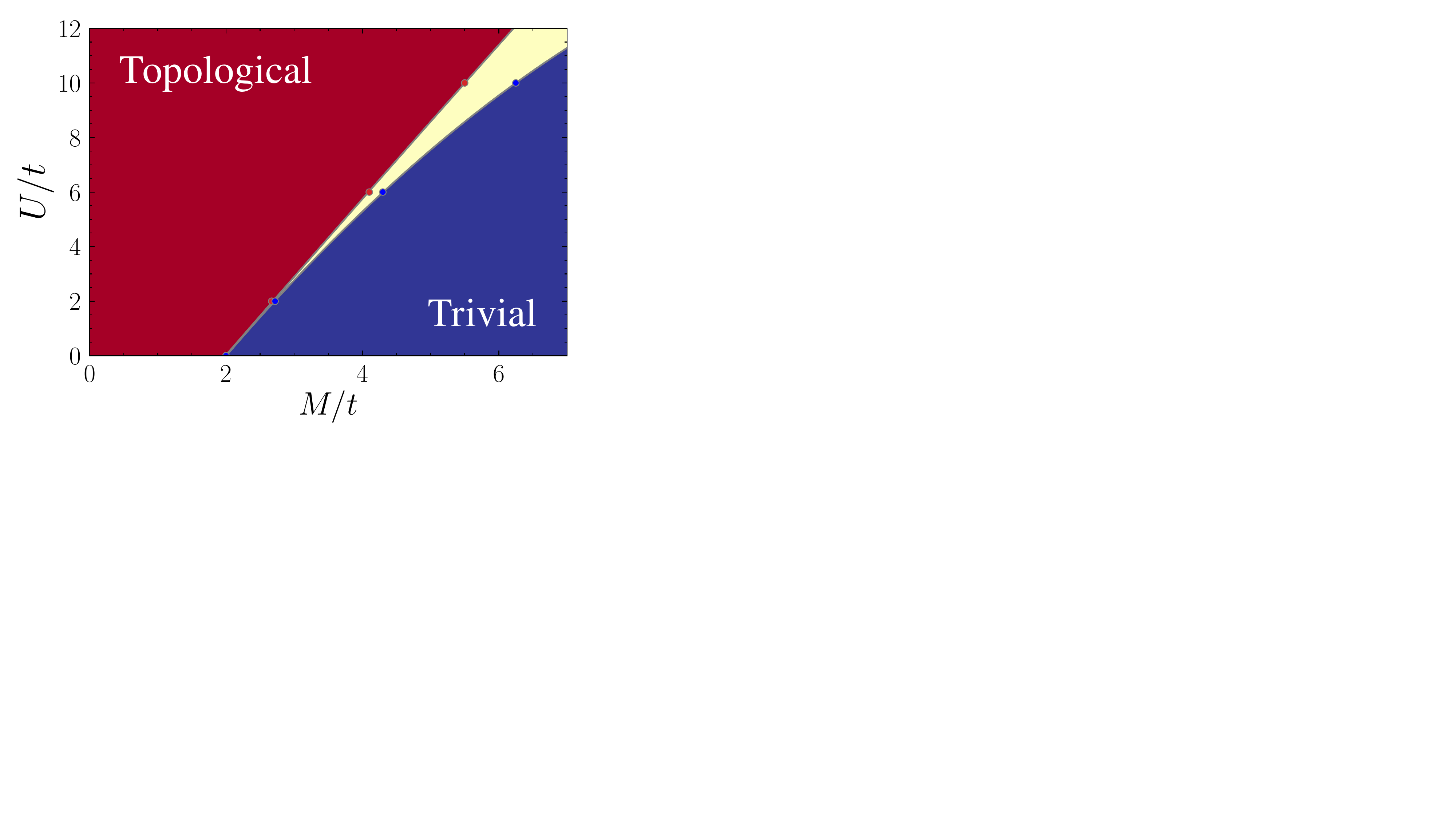}
\caption{\label{fig:pdV0} 
Phase diagram of the 1D BHZ model of Eq.~\eqref{eq:hamilt} in the $(M,U)$ plane for $V=0$ (and $t=\lambda$), as obtained using the Jastrow-Slater wave function. 
Topological (red) and trivial (blue) states are distinguished by the value of the phase of the many-body marker in Eq.~\eqref{eq:resta} ($0$ or $\pi$, respectively). 
For finite values of the intra-orbital interaction $U$, an intermediate (pale yellow) region is found, with complex values of the many-body marker. This insulating 
state sustains gapless spin excitations.}
\end{figure}

\begin{figure}
\includegraphics[width=\columnwidth]{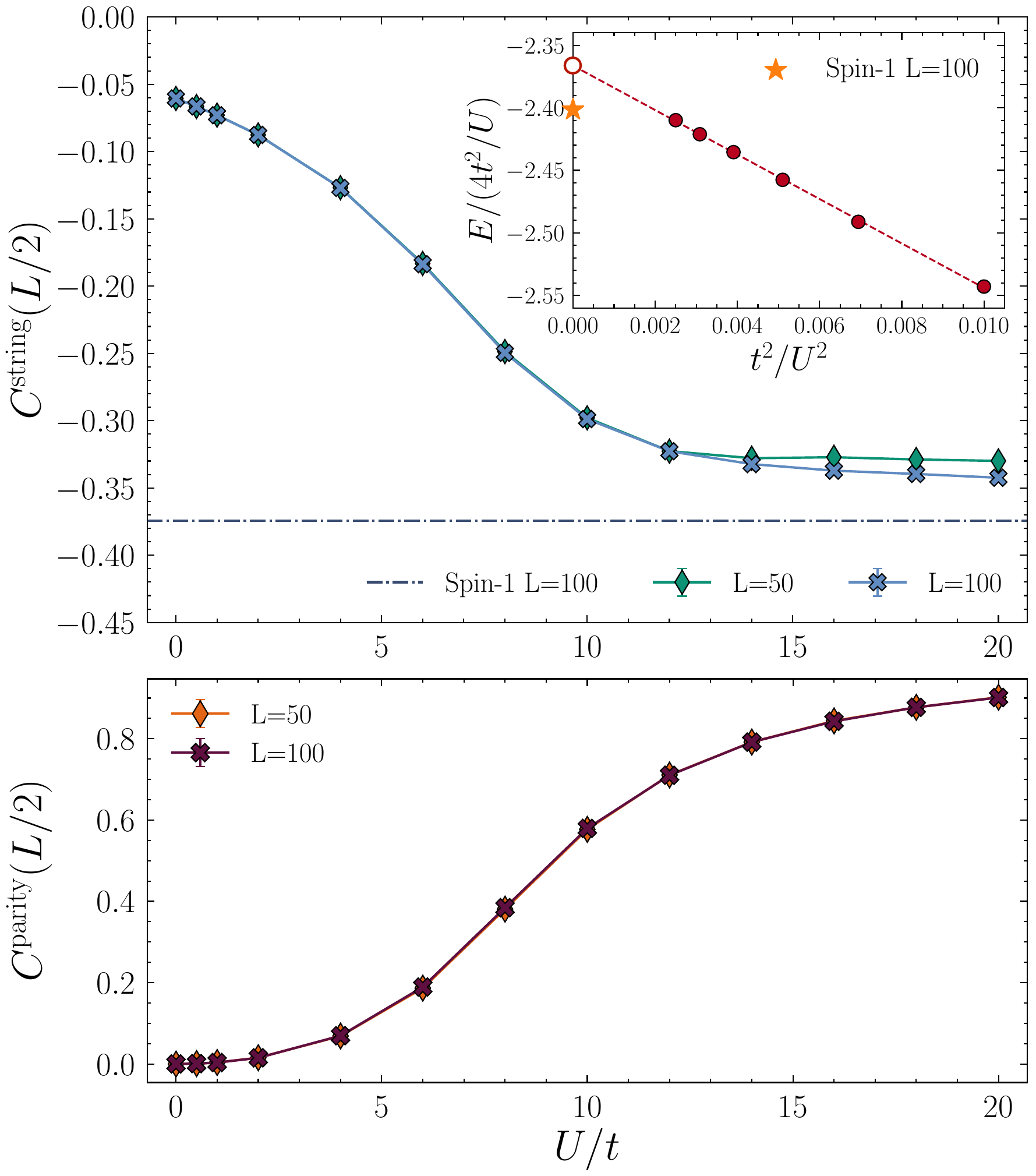}
\caption{\label{fig:string} 
Comparison between Jastrow-Slater results in the strong-coupling limit and DMRG calculations for the spin-1 Heisenberg model. Upper panel: String correlations in Eq.~\eqref{eq:string} at 
the maximum distance $C^{\mathrm{string}}(L/2)$ as a function of the intra-orbital interaction $U/t$, for $L=50$ and $L=100$ sites, at fixed on-site term $M/t=0.5$ 
and $V=0$. The value of the string-order parameter for the spin-1 Heisenberg model, obtained by DMRG calculations for system size $L=100$, is also reported (dashed 
line). The Jastrow-Slater energy per site (in unit of $4t^2/U$) as a function of $t^2/U^2$ is shown in the inset, compared with the DMRG one for $L=100$ spin-1 
Heisenberg model. Lower panel: The same as in the upper panel, but for the parity correlations in Eq.~\eqref{eq:parity} at the maximum distance $C^{\mathrm{parity}}(L/2)$, whose value for 
$U/t \to \infty$ approaches $1$.}
\end{figure}
%%%%%%%%%%%%%%%%%%%%%%%%%%%%%%%%%%%%%%%%%%%%%%%%%%%%%%%%%%%%%%%%%%%%%%%%%%%%%%%%%%%%%%%%%%%%%%%%%%%%%%%%%%%%%%%%%%%%%%%%%%%%%%%%%%%

\subsection{The case with $V=0$}

The ground-state phase diagram in the $(M,U)$ plane is shown in Fig.~\ref{fig:pdV0} and contains three different phases. The first one is a trivial insulator, which 
can be connected to both the non-interacting limit with $M/t>2$ and the Trivial state described within the atomic limit. The second one is a topological insulator, 
connected to the non-interacting state obtained for $M/t<2$ and the Uniform state described in the atomic limit. Unexpectedly, we find evidence for a third uniform 
phase, which intrudes in a narrow region between trivial and topological ones for $U>0$. Here our numerical calculations give evidence for gapless spin excitations 
at $q=0$ (while they are gapped at $q=\pi$), even though we cannot exclude the presence of a tiny gap. By contrast, density excitations are gapped, indicating an 
insulating character. In addition, this phase highlights the breaking of both $\eta$-PH and I+$\eta$ sign symmetries of Eqs.~\eqref{eq:PH} and~\eqref{eq:Inv}, while 
preserving the translational invariance. Let us describe the detailed calculations that allow us to determine the properties summarized in Fig.~\ref{fig:pdV0}. 

First, we investigate the effect of the intra-orbital interaction onto the topological state with $M/t<2$. Remarkably, by increasing the intra-orbital 
Hubbard-$U$ interaction (within the region $M/U \lesssim 0.5$), this state evolves continuously into the ground state of the spin-1 Heisenberg model, discussed in 
the strong-coupling regime in section~\ref{sec:atomic}. The latter one is characterized by having a non-vanishing (spin) string-order parameter~\cite{dennijs1989}. 
Within the electronic model, we can define the non-local correlation function:
\begin{equation}\label{eq:string}
C^{\mathrm{string}}(r) = \frac{1}{L} \sum_{j=1}^{L} \langle \hat{S}^z_{j} \exp{\left\{i\pi \sum_{l=j+1}^{j+r-1} \hat{S}^z_{l} \right\}} \hat{S}^z_{j+r} \rangle,
\end{equation}
where $\langle \dots \rangle$ indicates the expectation value over the approximated ground state, obtained by either variational Monte Carlo or DMRG approaches. 
Then the string-order parameter is given by the $r \to \infty$ limit of the correlation function. Our variational approach based on the Jastrow-Slater wave 
function of Eq.~\eqref{eq:wf} gives an accurate description of the strong-coupling limit. The results obtained for $M/t=0.5$ with increasing intra-orbital $U$ 
are presented in Fig.~\ref{fig:string}. Here the extrapolated energy per site (for $L=100$) is $E/J=-2.366(1)$, which compares well with the one of the spin-1 
model (including the constant term) $E/J=-2.4015$, obtained from the DMRG calculation (this latter result, computed for the $L=100$ chain with periodic boundary 
conditions, is numerically identical to the one extrapolated in the thermodynamic limit~\cite{white1993}). Note that $J=4t^2/U$ sets the correct energy scale of 
the strong-coupling limit with $V=0$, $M/U \to 0$, and $t=\lambda$, see section~\ref{sec:atomic}. Remarkably, the string-order parameter $C^{\mathrm{string}}(L/2)$ 
is finite not only in the strong-coupling regime but also down to $U=0$, as shown in Fig.~\ref{fig:string}. We mention that a finite value is also obtained in 
the non-interacting trivial phase. The fact that a string-order parameter may be finite in a non-interacting tight-binding model has been discussed in 
Ref.~\cite{anfuso2007}. In addition, we also evaluate the density parity-order parameter~\cite{montorsi2012,barbiero2013}, which provides information about 
the electron localization due to strong interactions. For that, we define the non-local correlation function:
\begin{equation}\label{eq:parity}
C^{\mathrm{parity}}(r) = \frac{1}{L} \sum_{j=1}^{L} \langle \exp{\left\{i\pi \sum_{l=j+1}^{j+r} \left (\hat{n}_{l} -2 \right )\right\}} \rangle.
\end{equation}
As before, the parity-order parameter is obtained from the $r \to \infty$ limit. In the strong-coupling limit, we have that $C^{\mathrm{parity}}(r)=1$, since each 
site is doubly occupied with no fluctuations. By contrast, in the non-interacting topological phase (i.e., $U=0$ and $M/t<2$) the parity-order parameter vanishes. 
The results shown in the lower panel in Fig.~\ref{fig:string} strongly indicate that $C^{\mathrm{parity}}(L/2)$ becomes finite as soon as the intra-orbital 
interaction is finite. As a consequence, the results on the string and parity correlations indicate that the non-interacting topological insulator is adiabatically 
connected to the strong-coupling limit, described by the spin-1 Heisenberg model. 

In the variational approach, the most relevant parameter necessary to correctly capture the physical properties of this phase is the fictitious antiferromagnetic 
field $h$ of Eq.~\eqref{eq:hamAF}, which becomes finite at $U>0$; the presence of the spin-spin Jastrow factor of Eq.~\eqref{eq:spinjas} is also crucial to prevent 
the instauration of a true magnetic ordering, similarly to what happens in the spin-1 Heisenberg chain~\cite{piccioni2023}.

%%%%%%%%%%%%%%%%%%%%%%%%%%%%%%%%%%%%%%%%%%%%%%%%%%%%%%%%%%%%%%%%%%%%%%%%%%%%%%%%%%%%%%%%%%%%%%%%%%%%%%%%%%%%%%%%%%%%%%%%%%%%%%%%%%%
\begin{figure}
\includegraphics[width=\columnwidth]{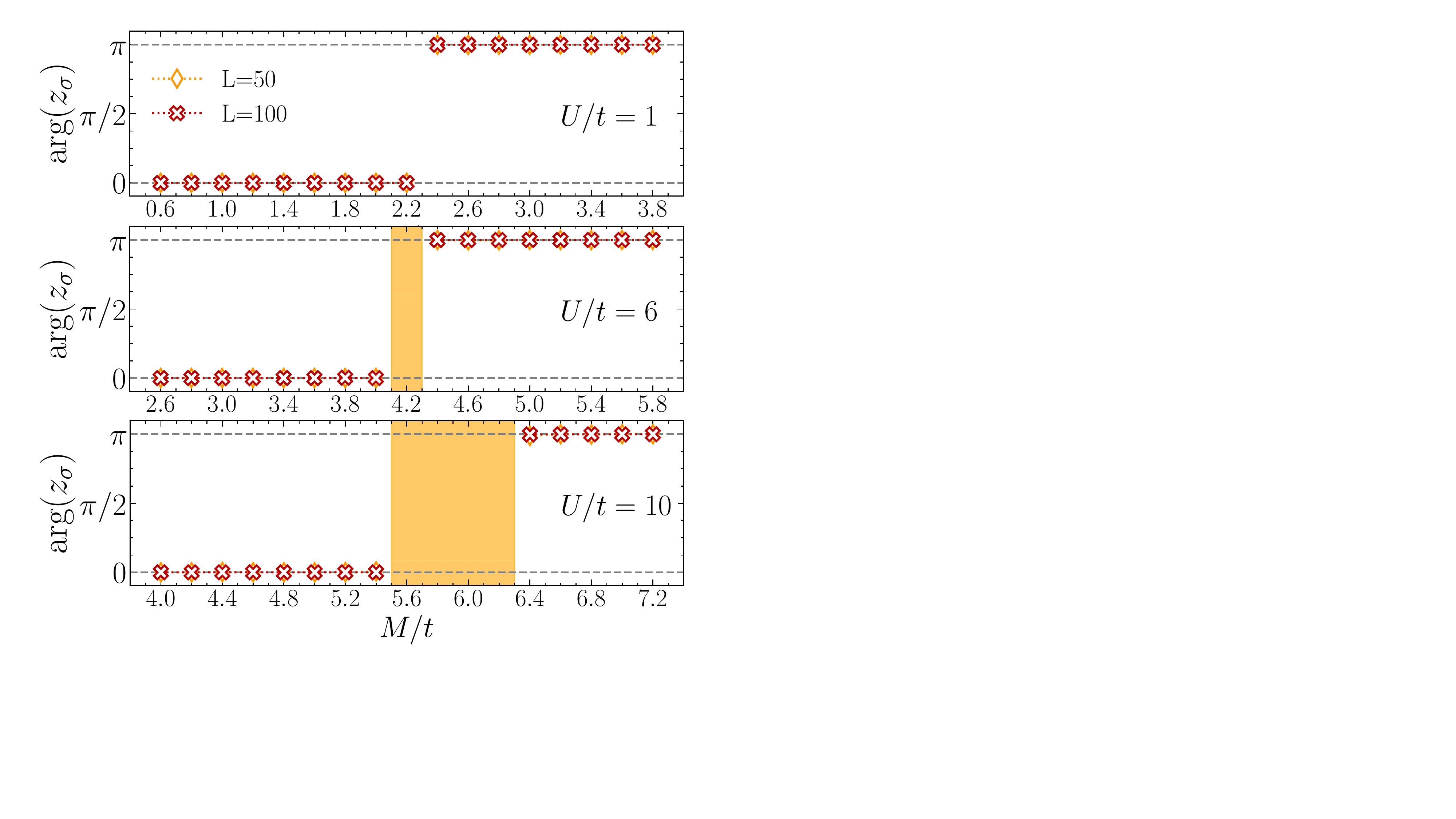}
\caption{\label{fig:zeta} 
The Jastrow-Salter results for the phase of many-body marker of Eq.~\eqref{eq:resta} at $V=0$ and $U/t=1$ (upper panel), $U/t=6$ (middle panel), and $U/t=10$ (lower 
panel). The yellow strips indicate regions where the phase of the many-body marker is not quantized.}
\end{figure}
%%%%%%%%%%%%%%%%%%%%%%%%%%%%%%%%%%%%%%%%%%%%%%%%%%%%%%%%%%%%%%%%%%%%%%%%%%%%%%%%%%%%%%%%%%%%%%%%%%%%%%%%%%%%%%%%%%%%%%%%%%%%%%%%%%%

Let us now move to describe the fate of the (continuous) transition between the topological and trivial phases when $U>0$. For that, we use the recently defined 
many-body marker in real space~\cite{gilardoni2022}, which builds on the pioneering work of Resta and Sorella for discriminating between metallic and insulating 
phases~\cite{resta1999}:
\begin{equation}\label{eq:resta}
z_{\sigma} = \langle \exp{\left\{i \frac{2 \pi}{L} \sum_{j=1}^{L} j \hat{n}_{j,\sigma} \right\}} \rangle.
\end{equation}
In absence of local order parameters distinguishing different phases of matter, this many-body marker provides valuable information about the nature of the many-body 
wave function~\cite{gilardoni2022}. In particular, for the non-interacting model, it is possible to show that $z_{\sigma}=(-1)^{L-1} e^{i \phi_{\sigma}}$, where 
$\phi_{\sigma}$ is the spin-resolved Zak phase; here, transition at $M/t=2$ can be detected by a jump in the phase of $z_{\sigma}$ (from $0$ to $\pi$). In general, 
for a fully gapped insulator that conserves the $z$ component of the total spin, the modulus of this quantity converges to $1$ in the thermodynamic limit. As long as either the $\eta$-PH or the I+$\eta$ sign symmetry [see Eqs.~\eqref{eq:PH} and~\eqref{eq:Inv}] is preserved, a {\it real} value of $z_{\sigma}$ is guaranteed, allowing 
us to discriminate between topological and trivial states according to its sign. By contrast, whenever the aforementioned symmetries are simultaneously broken, the 
many-body marker may become complex valued.

The phase of the many-body marker in Eq.~\eqref{eq:resta} is reported in Fig.~\ref{fig:zeta} for three values of $U$ for chains with $L=50$ and $100$ sites. For $U/t=1$, 
$z_{\sigma}$ is always real and its sign has an abrupt change from $M/t=2.2$ to $2.4$, even if we cannot exclude the presence of a very narrow intermediate phase by 
refining the grid of points on which calculating the marker. The situation noticeably changes increasing the value of the intra-orbital interaction $U$. For $U/t=6$ and 
$U/t=10$, there are values of $M/t$ for which $z_{\sigma}$ is complex. For example, at $U/t=10$, this region extends from $M/t \approx 5.5$ to $6.1$. We mention that 
a similar outcome also appears within DMRG calculations, where the intermediate region with a complex $z_{\sigma}$ is detected for $5.3 \lesssim M/t \lesssim 5.9$.

Within this intermediate region, the variational state is defined by an auxiliary Hamiltonian~\eqref{eq:hamAF} in which the intra-orbital parameters $\tilde{M}$ 
and $\tilde{t}_{k}$ converge to values that would give rise to an almost gapless spectrum; however, the on-site inter-orbital hopping parameter, taken as 
$\tilde{\lambda}_{0,\sigma} = i \sigma \lambda_0$ with a real $\lambda_0$, becomes finite (as well as a nearest-neighbor inter-orbital term 
$\tilde{\lambda}_{1,\eta,\sigma}=i \eta s_{\sigma} \gamma_{1}$, with a real $\gamma_{1}$), thus opening a (small) energy gap in the non-interacting spectrum. 
These inter-orbital terms explicitly break the $\eta$-PH and I+$\eta$ sign symmetries (see Appendix~\ref{sec:app}), as highlighted by the complex value of the 
many-body marker in Fig.~\ref{fig:zeta}. The variational state preserves translational invariance, since the optimal value of the CDW parameters $\Delta$ is 
vanishingly small.

%%%%%%%%%%%%%%%%%%%%%%%%%%%%%%%%%%%%%%%%%%%%%%%%%%%%%%%%%%%%%%%%%%%%%%%%%%%%%%%%%%%%%%%%%%%%%%%%%%%%%%%%%%%%%%%%%%%%%%%%%%%%%%%%%%%
\begin{figure}
\includegraphics[width=\columnwidth]{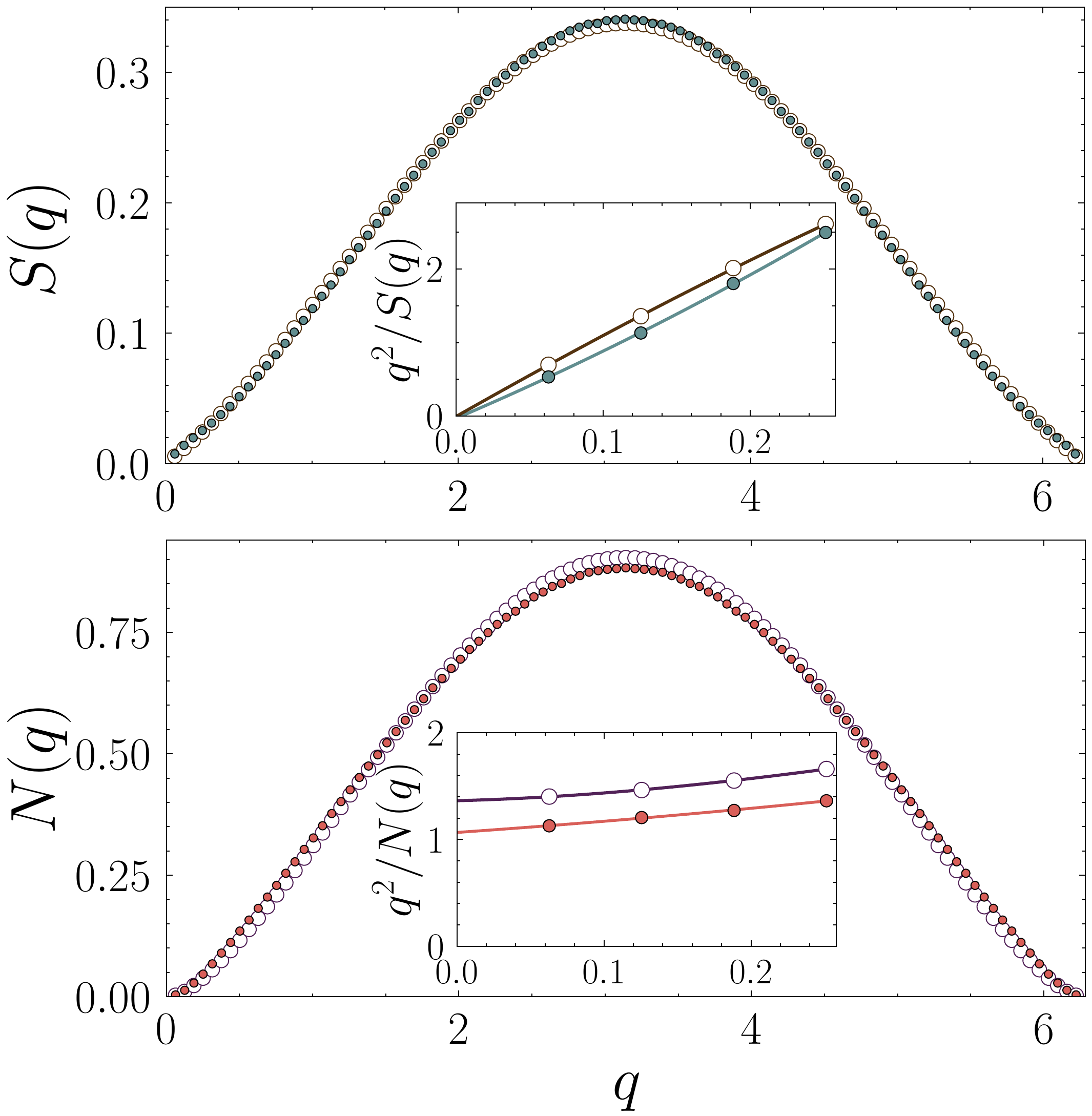}
\caption{\label{fig:sqnqinno} 
Spin (upper panel) and density (lower panel) structure factors in the intermediate phase (between topological and trivial insulators), for $L=100$ sites at $V=0$, 
$U/t=10$, and $M/t=5.9$, obtained by the Jastrow-Slater wave functions (full points). The DMRG results are also reported for comparison (empty points). The insets 
show the ratios $q^2/S(q)$ and $q^2/N(q)$ for small momenta, in order to emphasize the linear or quadratic behavior of the structure factors.}
\end{figure}
%%%%%%%%%%%%%%%%%%%%%%%%%%%%%%%%%%%%%%%%%%%%%%%%%%%%%%%%%%%%%%%%%%%%%%%%%%%%%%%%%%%%%%%%%%%%%%%%%%%%%%%%%%%%%%%%%%%%%%%%%%%%%%%%%%%

In order to clarify the nature of the intermediate phase and assess the existence of gapless excitations, we consider the density and spin structure factors:
\begin{eqnarray}\label{eq:nq}
N(q) &=& \frac{1}{L} \sum_{j=1}^{L} e^{i q r} \langle \hat{n}_{j} \hat{n}_{j+r} \rangle, \\
S(q) &=& \frac{1}{L} \sum_{j=1}^{L} e^{i q r} \langle \hat{S}^{z}_{j} \hat{S}^{z}_{j+r} \rangle, 
\label{eq:sq}
\end{eqnarray}
where $\hat{n}_{j}$ and $\hat{S}^{z}_{j}$ are the total density and spin operators on the $j$ site, respectively. In fact, gapless excitations at $q=0$ are marked by 
a linear behavior of these quantities for $q \to 0$~\cite{feynman1956,capello2005}; a diverging peak at a finite $q$ would also imply a gapless mode at that momentum. 
A standard example is given by the spin-1/2 Heisenberg chain, where $S(q) \propto q$ at small momenta and $S(\pi) \propto \ln(L)$. For both topological and trivial 
phases shown in the phase diagram of Fig.~\ref{fig:pdV0}, $N(q)$ and $S(q)$ do not possess any singularity, with a quadratic behavior for small values of the momenta 
(not shown), in agreement with the fact that both density and spin excitations are gapped, as in the non-interacting limit. By contrast, a different scenario appears 
within the intermediate phase, where the many-body marker is not quantized, see Fig.~\ref{fig:sqnqinno}. While the density structure factor preserves a completely 
smooth behavior, indicating an insulating character of the state, the spin structure factor has a linear behavior at small momenta, e.g., $S(q) \propto q$, signaling 
the existence of {\it gapless} excitations. Nevertheless, the situation is strikingly different from the standard 1D Heisenberg model, since there is neither a 
diverging peak nor a kink at $q=\pi$. All these features are reproduced by DMRG calculations. It should be mentioned that wave functions with a metallic character 
[i.e., with both $S(q)$ and $N(q)$ linear in $q$ for small momenta] may be also obtained but having a slightly higher variational energy.
 
%%%%%%%%%%%%%%%%%%%%%%%%%%%%%%%%%%%%%%%%%%%%%%%%%%%%%%%%%%%%%%%%%%%%%%%%%%%%%%%%%%%%%%%%%%%%%%%%%%%%%%%%%%%%%%%%%%%%%%%%%%%%%%%%%%%
\begin{figure}
\includegraphics[width=\columnwidth]{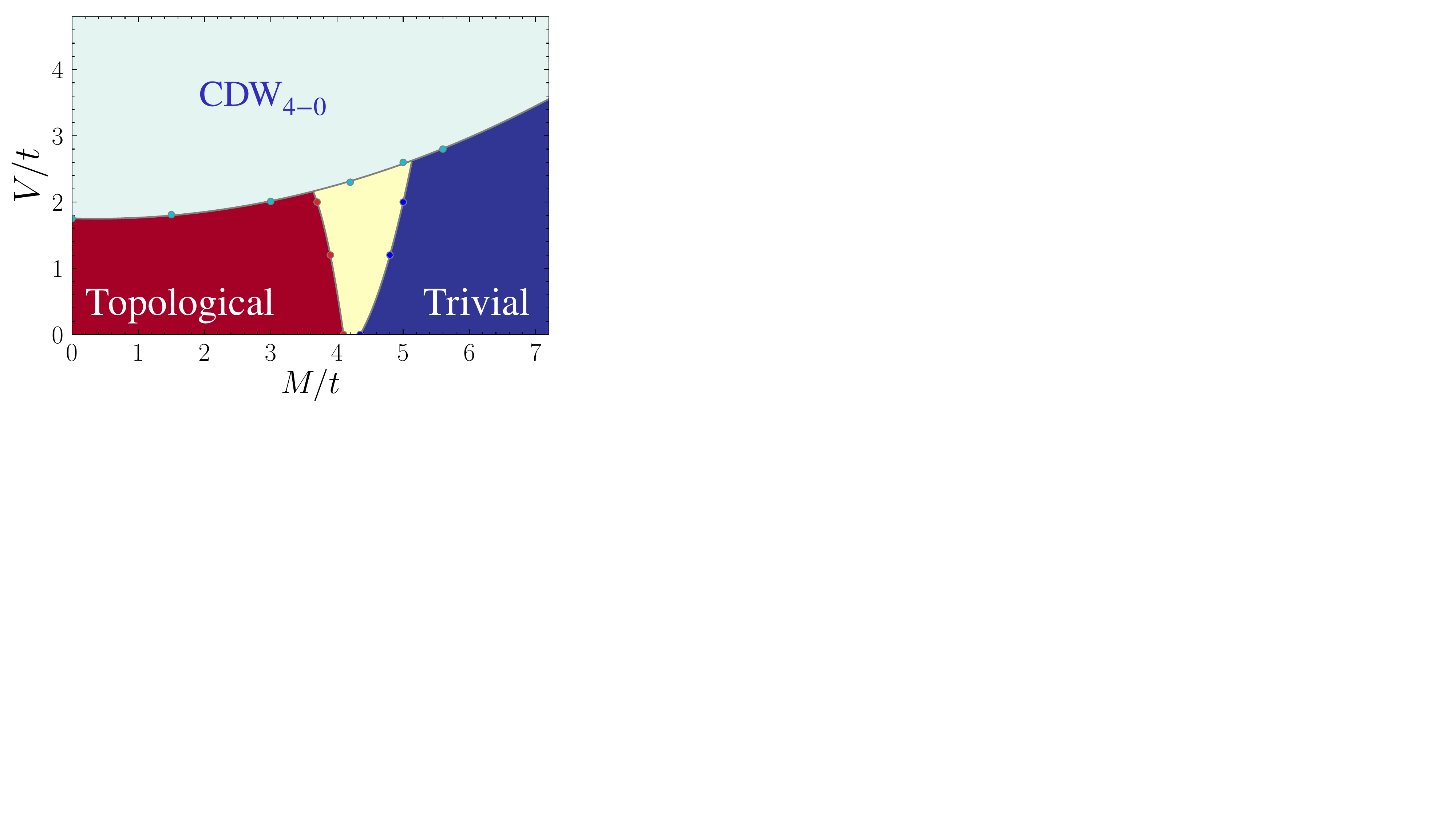}
\caption{\label{fig:pdU6}
Phase diagram of the 1D BHZ model of Eq.~\eqref{eq:hamilt} in the $(M,V)$ plane for $U/t=6$ (and $t=\lambda$), as obtained using the Jastrow-Slater wave function. 
Besides the topological (red), trivial (blue), and intermediate (pale yellow) phases, observed in the case with $V=0$, as shown in Fig.~\ref{fig:pdV0}, a fully 
gapped CDW (pale blue) exits, for large-enough values of $V/t$.}
\end{figure}

\begin{figure}
\includegraphics[width=\columnwidth]{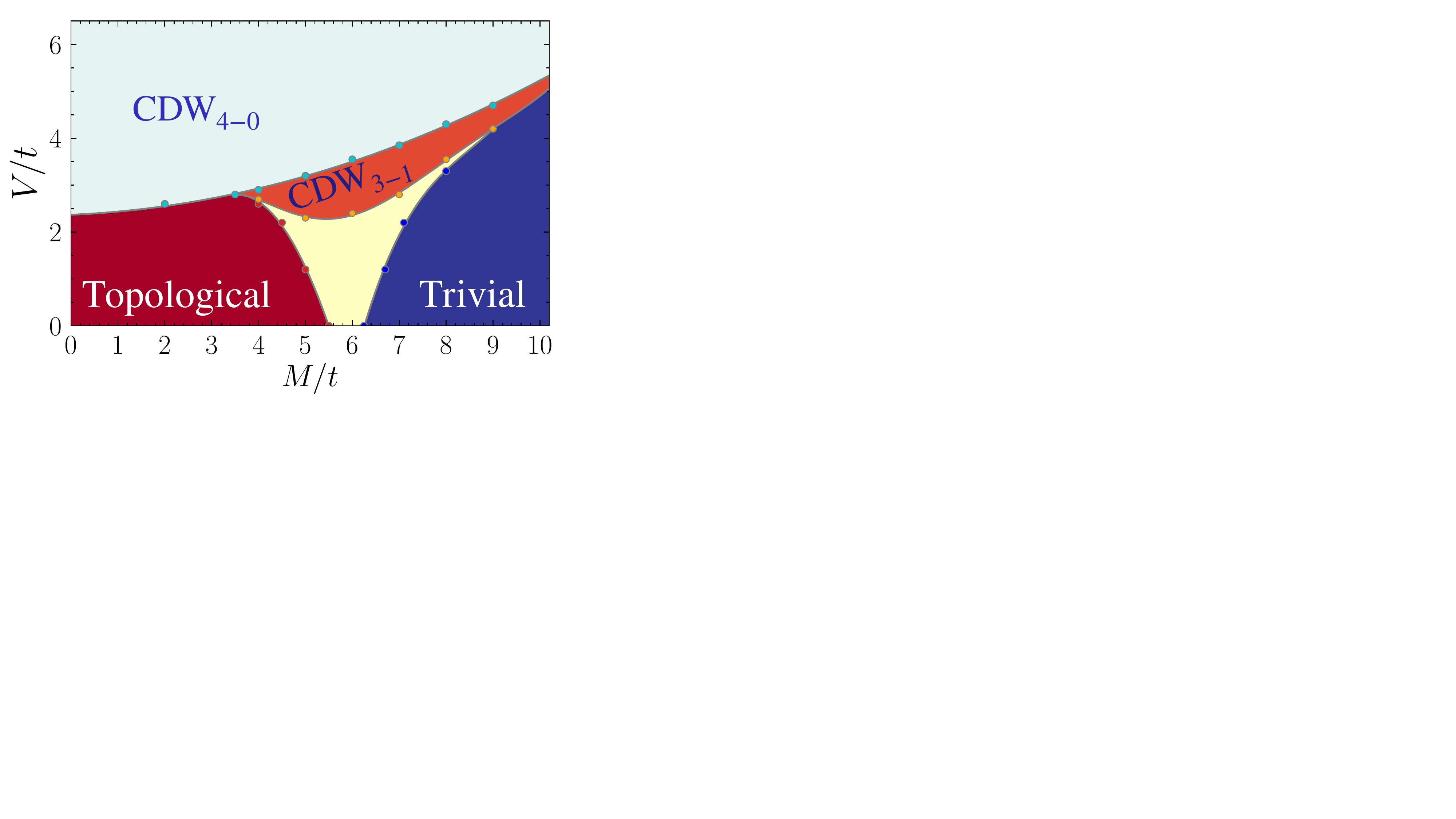}
\caption{\label{fig:pdU10}
Phase diagram of the 1D BHZ model of Eq.~\eqref{eq:hamilt} in the $(M,V)$ plane for $U/t=10$ (and $t=\lambda$), as obtained using the Jastrow-Slater wave function. 
In addition to the phases shown in Fig.~\ref{fig:pdU6}, i.e., topological (red), trivial (blue), intermediate (pale yellow), and CDW$_{4-0}$ (pale blue), a 
CDW$_{3-1}$ with gapless spin excitations (orange) is stabilized, between the intermediate and the fully gapped CDW phases.}
\end{figure}
%%%%%%%%%%%%%%%%%%%%%%%%%%%%%%%%%%%%%%%%%%%%%%%%%%%%%%%%%%%%%%%%%%%%%%%%%%%%%%%%%%%%%%%%%%%%%%%%%%%%%%%%%%%%%%%%%%%%%%%%%%%%%%%%%%%

\subsection{The case with $V>0$}

Let us move to the case with $V>0$. For that, we consider two values of the intra-orbital interaction $U/t=6$ and $U/t=10$ and show the phase diagram in the $(M,V)$ 
plane, see Figs.~\ref{fig:pdU6} and~\ref{fig:pdU10}. The first observation is that a (not too large) nearest-neighbor repulsion $V$ enlarges the stability region of 
the intermediate phase. Most importantly, two additional phases settle down. The first one is a fully gapped CDW, which appears for sufficiently large values of $V$; 
it breaks the translational symmetry by one lattice spacing and is adiabatically connected to the CDW$_{4-0}$ that has been discussed in section~\ref{sec:atomic}. 
The second one is an insulating CDW, still breaking the translational symmetry but sustaining gapless spin excitations, connected to the CDW$_{3-1}$ presented in 
section~\ref{sec:atomic}. Although the local densities are non-integer for finite values of $U$ and $V$, we refer to these insulators as CDW$_{4-0}$ and CDW$_{3-1}$. 
In all cases, the presence of a density disproportionation is marked by the (local) order parameter that is computed from density-density correlations in real space:
\begin{equation}\label{eq:cdworder}
m_{\mathrm{cdw}} = \lim_{L \to \infty} |\tilde{N}(L/2)-\tilde{N}(L/2-1)|,
\end{equation}
where
\begin{equation}
\tilde{N}(r) = \frac{1}{L} \sum_{j=1}^{L} \langle \hat{n}_{j} \hat{n}_{j+r} \rangle.
\end{equation}
In the atomic limit, the values of the local parameter for the CDW$_{4-0}$ and CDW$_{3-1}$ are $m_{\mathrm{cdw}}=8$ and $2$, respectively. 

%%%%%%%%%%%%%%%%%%%%%%%%%%%%%%%%%%%%%%%%%%%%%%%%%%%%%%%%%%%%%%%%%%%%%%%%%%%%%%%%%%%%%%%%%%%%%%%%%%%%%%%%%%%%%%%%%%%%%%%%%%%%%%%%%%%
\begin{figure}
\includegraphics[width=\columnwidth]{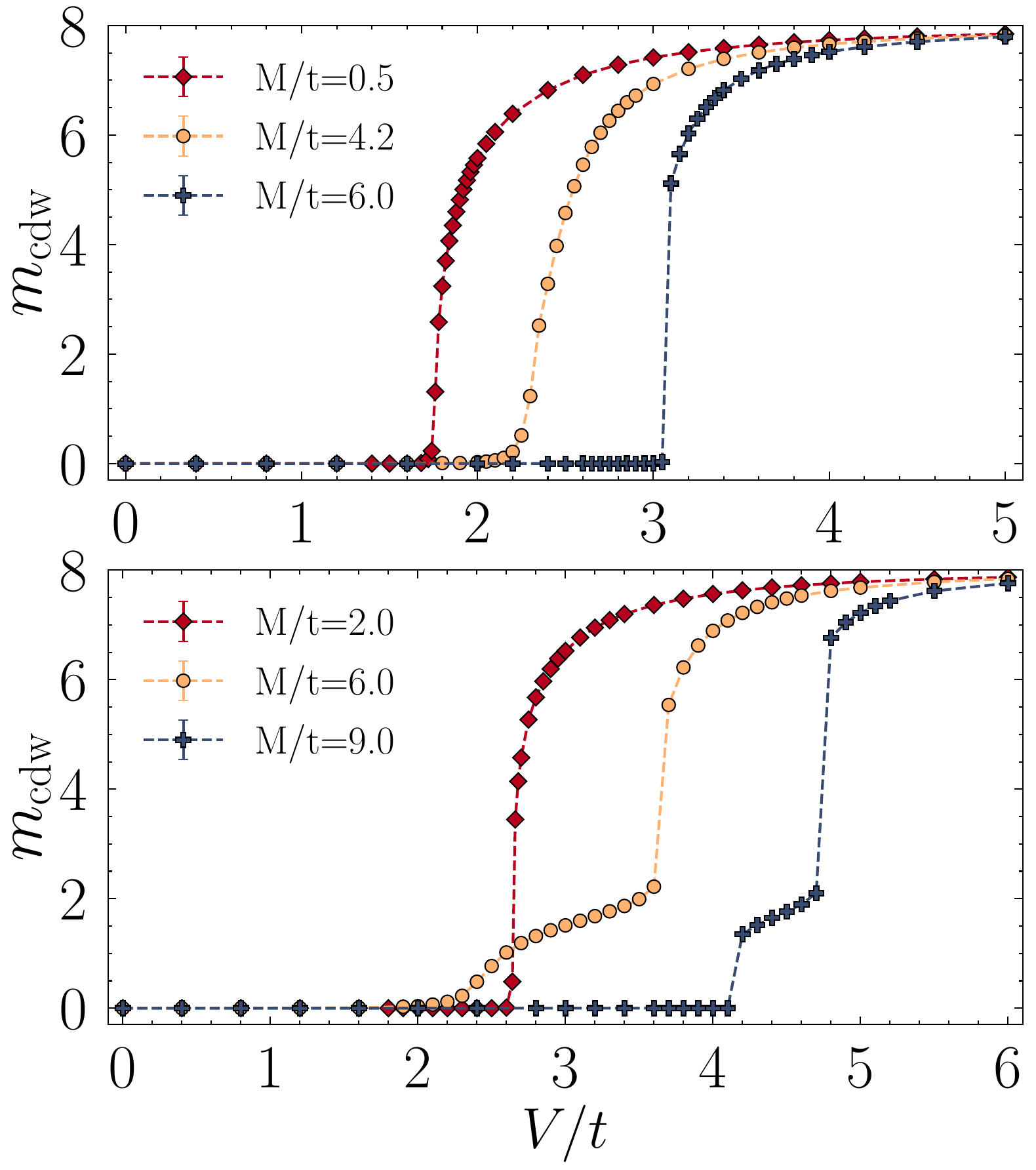}
\caption{\label{fig:mCDW}
The Jastrow-Salter results for the CDW order parameter $m_{\mathrm{cdw}}$ in Eq.~\eqref{eq:cdworder} as a function of $V/t$ for $L=50$ sites at $U/t=6$ (upper panel) 
and $U/t=10$ (lower panel).}
\end{figure}
%%%%%%%%%%%%%%%%%%%%%%%%%%%%%%%%%%%%%%%%%%%%%%%%%%%%%%%%%%%%%%%%%%%%%%%%%%%%%%%%%%%%%%%%%%%%%%%%%%%%%%%%%%%%%%%%%%%%%%%%%%%%%%%%%%%

For the values of $U/t$ considered here, the transition between the topological insulator and the fully gapped CDW is continuous, see Fig.~\ref{fig:mCDW}. By contrast, 
a first-order transition (with a jump in $m_{\mathrm{cdw}}$) is clearly detected when the intra-orbital Hubbard interaction is increased further, e.g., $U/t=14$ (not 
shown). This result is in agreement with DMRG calculations (on chains with open-boundary conditions) of Ref.~\cite{barbarino2019}. 

\begin{figure}
\includegraphics[width=\columnwidth]{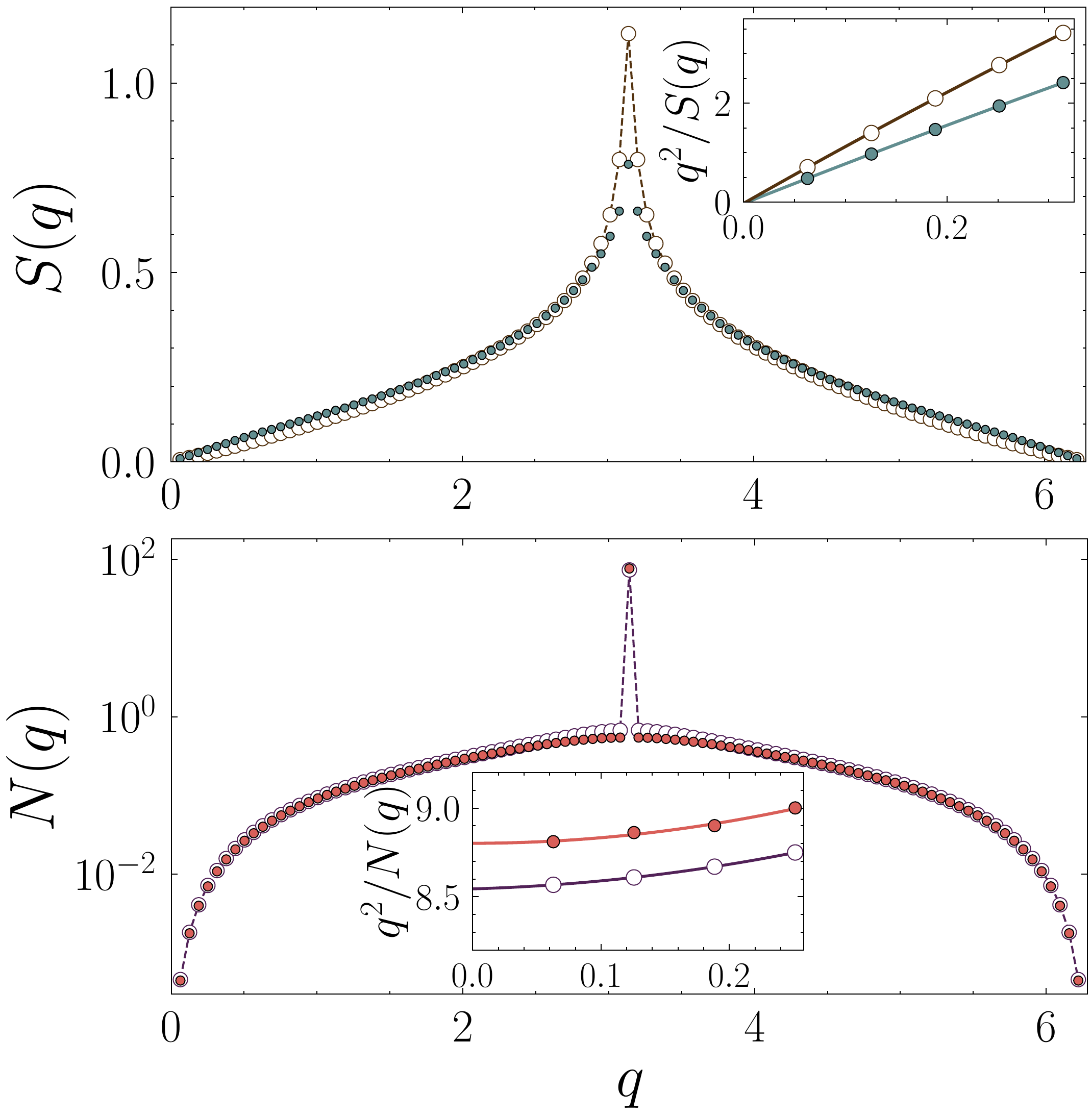}
\caption{\label{fig:sqnq31}
The same in Fig.~\ref{fig:sqnqinno} for the CDW$_{3-1}$ phase, with $L=100$ sites at $V/t=3$, $U/t=10$ and $M/t=6$. Notice the logarithmic scale in the 
$y$ axis in the lower panel, taken for showing the huge value of $N(\pi)$.}
\end{figure}

\begin{figure}
\includegraphics[width=\columnwidth]{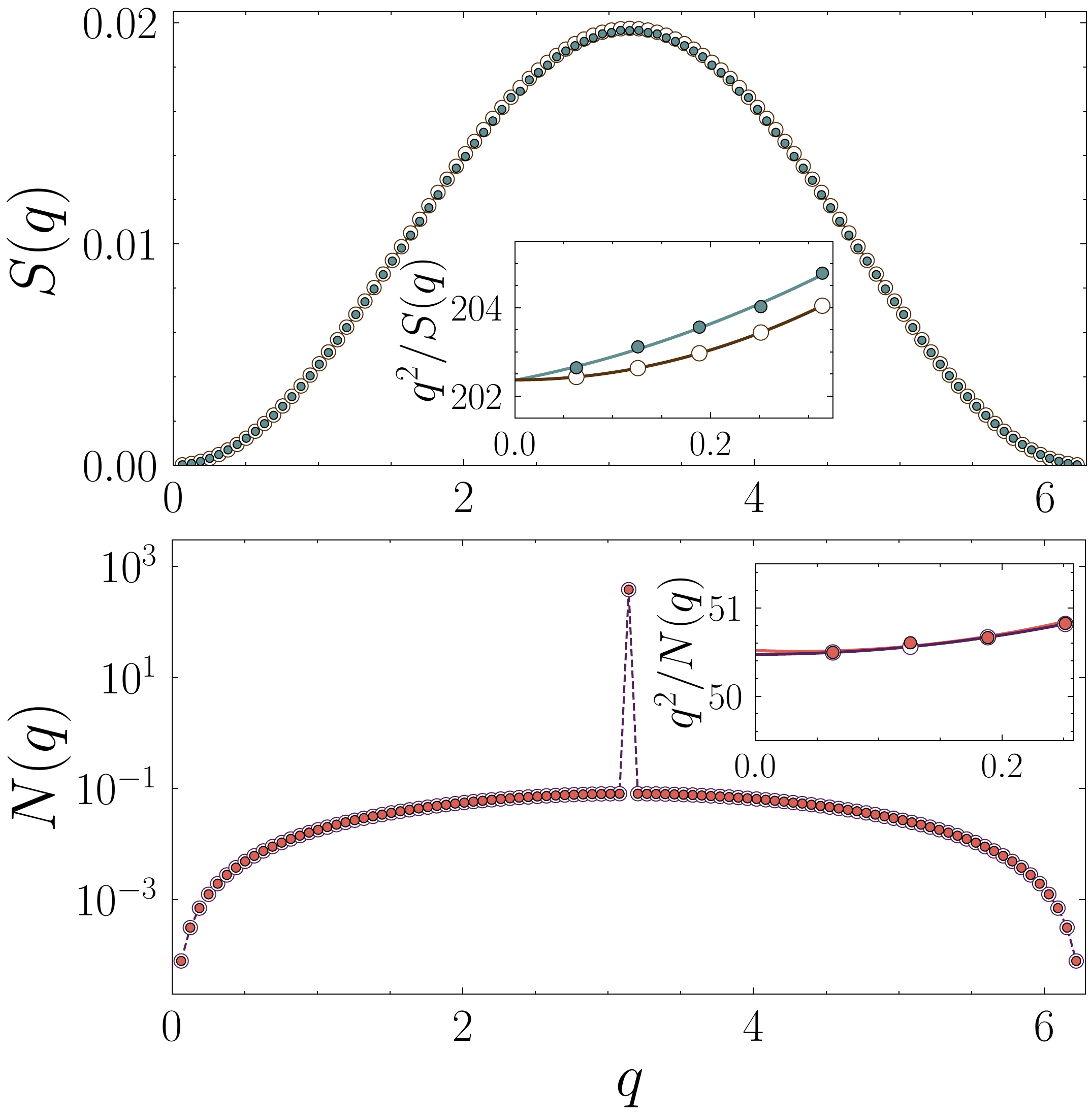}
\caption{\label{fig:sqnq40}
The same as in Fig.~\ref{fig:sqnq31} for the CDW$_{4-0}$ phase, with $L=100$ sites at $V/t=5$, $U/t=10$ and $M/t=6$. Notice the logarithmic scale in the 
$y$ axis in the lower panel, taken for showing the huge value of $N(\pi)$.}
\end{figure}

For $U/t=6$, the intermediate region (where the many-body marker is complex) is characterized by a vanishingly small CDW order parameter up to moderate values of 
$V$. For this value of the on-site repulsion, no sign of the CDW$_{3-1}$ is detected. Then, by increasing further the nearest-neighbor interaction, a continuous 
transition to a fully gapped CDW is observed; see Fig.~\ref{fig:mCDW}. Furthermore, the transition between the trivial and the CDW$_{4-0}$ insulator is most likely 
first-order (see Fig.~\ref{fig:mCDW}). The situation changes for $U/t=10$, where the CDW$_{3-1}$ phase appears in a rather wide region of the phase diagram, in 
between the intermediate insulating phase and the CDW$_{4-0}$ one; see Fig.~\ref{fig:pdU10}. The presence of the CDW$_{3-1}$ phase can be detected from 
$m_{\mathrm{cdw}}$, which shows a ``step-like'' behavior when increasing $V$, as shown in Fig.~\ref{fig:mCDW}. Most importantly, the spin structure factor $S(q)$ 
acquires singularities at $q=0$ and $\pi$, thus signaling the existence of gapless modes, see Fig.~\ref{fig:sqnq31}. The presence of a huge peak in the density 
structure factor at $q=\pi$ is the hallmark of the insurgence of charge disproportionation [i.e., $N(\pi) \approx m_{\mathrm{cdw}} L$], see Fig.~\ref{fig:sqnq31}. 
From our variational calculations, the transition between the intermediate state and the CDW$_{3-1}$ one is most likely continuous, as deduced from the smooth 
behavior of $m_{\mathrm{cdw}}$ in Fig.~\ref{fig:mCDW}; instead, the transition from gapless to gapped CDW phases looks first-order, with a sudden jump in 
$m_{\mathrm{cdw}}$ (from our data, however, we cannot exclude a second-order transition, with a small critical exponent). Remarkably, within the CDW$_{3-1}$ the 
value of $m_{\mathrm{cdw}}$ is always smaller than the ``saturation'' limit (i.e., $m_{\mathrm{cdw}}=2$). We emphasize that the latter quantity cannot be used 
to distinguish between CDW$_{3-1}$ and CDW$_{4-0}$, which also show qualitatively similar (smooth) density-density structure factors $N(q)$, compare 
Figs.~\ref{fig:sqnq31} and~\ref{fig:sqnq40}. By contrast, these two CDW phases can be distinguished by the presence of singularities in $S(q)$: while the fully 
gapped CDW$_{4-0}$ has a smooth spin-spin structure factor, the gapless CDW$_{3-1}$ displays singularities for both $q=0$ and $q=\pi$, compare Figs.~\ref{fig:sqnq31} 
and~\ref{fig:sqnq40}.

We conclude by mentioning that in both CDW$_{3-1}$ and CDW$_{4-0}$ regions, the optimized variational wave function is constructed from the auxiliary Hamiltonian 
containing, besides on-site and hopping parameters, a sizable CDW parameter $\Delta$, see Eq.~\eqref{eq:hamCDW}. In addition, the quasi-long range order in the spin 
sector is triggered by the fictitious magnetic field $h$ in Eq.~\eqref{eq:hamAF} and the spin-spin Jastrow factor in Eq.~\eqref{eq:spinjas}~\cite{franjic1997}.

%%%%%%%%%%%%%%%%%%%%%%%%%%%%%%%%%%%%%%%%%%%%%%%%%%%%%%%%%%%%%%%%%%%%%%%%%%%%%%%%%%%%%%%%%%%%%%%%%%%%%%%%%%%%%%%%%%%%%%%%%%%%%%%%%%%
%%%%%%%%%%%%%%%%%%%%%%%%%%%%%%%%%%%%%%%%%%%%%%%%%%%%%%%%%%%%%%%%%%%%%%%%%%%%%%%%%%%%%%%%%%%%%%%%%%%%%%%%%%%%%%%%%%%%%%%%%%%%%%%%%%%

\section{Conclusions}\label{sec:concl}

In this work, we studied a 1D electronic model that, in the non-interacting limit, features topological and trivial insulating phases, separated by a continuous 
transition, where the band gap closes. By including electron-electron repulsion, a rich phase diagram is obtained. In presence of the on-site (intra-orbital) 
Hubbard-$U$, the topological insulator is adiabatically connected to the Haldane (spin-1 gapped) insulator. Interestingly, the transition from the trivial to 
the topological phase is no longer direct, but an intermediate phase intrudes in a small region between them. This intermediate phase (still insulating) is 
characterized by gapless spin excitations at $q=0$ and a complex many-body marker~\eqref{eq:resta}, which indicates that both $\eta$-PH and I+$\eta$ sign 
symmetries of Eqs.~\eqref{eq:PH} and~\eqref{eq:Inv} are broken. By adding the nearest-neighbor $V$ interaction~\cite{barbarino2019}, two additional phases 
emerge, breaking translational symmetries. One of these phases (CDW$_{4-0}$) is stabilized for large-enough values of $V$ and is fully gapped. The other one 
(CDW$_{3-1}$) appears when $U$ and $V$ have comparable values and sustains gapless spin excitations at both $q=0$ and $q=\pi$. Among all the phases obtained here, 
the intermediate one does not have a simple representation in the weak- or strong-coupling limit. Still, our numerical calculations, based on both Jastrow-Slater 
variational wave functions and DMRG approaches, strongly suggest that the intermediate gapless phase is a genuine state of the system. Further investigations, 
potentially including a weak-coupling bosonization approach~\cite{giamarchibook} starting from the gapless non-interacting point $M/t=2$ or the definition of 
alternative topological markers, will help confirm the existence of this intriguing phase of matter. 

\section*{Acknowledgements}
We acknowledge the CINECA award under the ISCRA initiative for the availability of high performance computing resources and support. We thank M. Fabrizio and 
T. Giamarchi for useful discussions. R.F. and D.P. also thank I. Pasqua for explaining some aspects of related works and L. L. Viteritti for carefully reading this 
paper.

\appendix
\section{Symmetries of the auxiliary Hamiltonian}\label{sec:app}

Here we briefly sketch the effect of time-reversal and lattice symmetries on the parameters that enter the auxiliary Hamiltonian in Eq.~\eqref{eq:auxiliary},
used to define the variational wave function. Indeed, while the intra-orbital hoppings are chosen to respect all the symmetries of the physical Hamiltonian, 
possible (spontaneous) breaking of these symmetries are encoded into the inter-orbital hoppings and/or the fictitious magnetic field $h$ and CDW $\Delta$ parameters, 
see Eq.~\eqref{eq:hamAF} and Eq.~\eqref{eq:hamCDW}, respectively. We have that:
 
\begin{equation}
\text{TR} \implies \begin{cases}
 \tilde{\lambda}_{0,\uparrow} = \tilde{\lambda}^{*}_{0,\downarrow}, \\
 \tilde{\lambda}_{k,\eta,\uparrow} = \tilde{\lambda}^{*}_{k,\eta,\downarrow},
\end{cases}
\end{equation}

\begin{equation}
 \text{SF}+\eta~\text{sign} \implies \begin{cases}
 \tilde{\lambda}_{0,\uparrow} = -\tilde{\lambda}_{0,\downarrow}, \\
 \tilde{\lambda}_{k,\eta,\uparrow} = -\tilde{\lambda}_{k,\eta,\downarrow},
\end{cases}
\end{equation}

\begin{equation}
\eta\text{-PH} \implies \begin{cases}
 \tilde{\lambda}_{0,\sigma} = -\tilde{\lambda}_{0,\sigma}, \\
 \tilde{\lambda}_{k,\eta,\sigma} = -\tilde{\lambda}^{*}_{k,-\eta,\sigma}.
\end{cases}
\end{equation}

\begin{equation}
\text{I}+\eta~\text{sign} \implies \begin{cases}
 \tilde{\lambda}_{0,\sigma} = -\tilde{\lambda}_{0,\sigma}, \\
 \tilde{\lambda}_{k,\eta,\sigma} = -\tilde{\lambda}^{*}_{k,-\eta,\sigma}.
\end{cases}
\end{equation}
When preserving all these symmetry, we obtain that $\tilde{\lambda}_{0,\sigma}=0$ and $\tilde{\lambda}_{k,\eta,\sigma}=i s_{\sigma} \lambda_{k}$ 
(with $\lambda_{k}$ real), as in the 1D BHZ Hamiltonian in Eq.~\eqref{eq:h0} (which only contains the $k=1$ term).

Finally, when the projector $\mathcal{P}_{\{S^{z}=0\}}$ is present, as shown in Eq.~\eqref{eq:wf}, the $h$ term does not break any of the three local symmetries 
(TR, SF+$\eta$ sign, and $\eta$-PH) or the lattice symmetries (I+$\eta$ sign and the translational one). On the other hand, the CDW term explicitly breaks 
$\eta$-PH and T, while preserving all the others.

\bibliography{biblio} 
\end{document}